\documentstyle[prl,aps,preprint]{revtex}
\begin{document}
\draft
\preprint{\vbox{To be published in {\it Physical Review C}
\hfill La Plata-TH-95-24}\\}

\title
{Temperature and Density Effects on the Nucleon Mass
Splitting\thanks{The authors are partially supported by 
CONICET, Argentina.}}
\author{
H. R. Christiansen, L. N. Epele, H. Fanchiotti \\ and
C. A. Garc\'\i a Canal
\\ {\normalsize\it Departamento de F\'\i sica, Universidad
Nacional de La Plata}\\ {\normalsize\it C.C. 67, (1900) La Plata,
Argentina}}

\maketitle

\begin{abstract}
The finite temperature and finite density dependence
of the neu\-tron-proton mass difference is analysed
in a purely hadronic framework where  the $\rho-\omega $ mixing is
crucial for this isospin symmetry breakdown.
The problem is handled within Thermo Field Dynamics. The present
results,  consistent with  partial chiral and charge symmetry restoration,
improve the experimental data fit for the energy difference between
mirror nuclei.
\end{abstract}
\vskip 1cm
\newpage
\pagenumbering{arabic}

\section{Introduction}

Symmetry restoration in spontaneously broken gauge
theories has been one of the first applications of field theory at
finite temperature \cite{ft}.
In the Standard Model, such a phase transition is supposed to have
taken place in the early Universe
when the temperature was of the order of a few hundred GeV.
Different types of phase transitions are also
expected to occur  at much lower
temperatures and with hadronic non zero densities.
In connection with it, the broken chiral symmetry has been predicted
to be restored at about the same temperature at which the deconfinement
sets in. This temperature is in the range of 150 to 200 MeV
as it is indicated
from lattice calculations \cite{latt} and  phenomenological models
based on effective Lagrangians \cite{eff}. High density
conditions are supposed to have  similar consequences on the
hadronic matter.

In the nuclear scale there are also indications of partial deconfinement
in order to explain the EMC effect \cite{emc}. It has been argued that
nucleons in nuclei occupy a larger volume than nucleons in vacuum
\cite{swel1}, a phenomenom that could be understood as a nucleon
swelling \cite{swel2}.
It has been also analysed in connection with the Nolen Schiffer Anomaly
(NSA) \cite{nsa} for different models of the nucleon
\cite{gonz}-\cite{krein}.

In recent years, the possibility of producing quark-gluon plasma by means
of very
energetic nucleus-nucleus collisions \cite{qmat} opened a rather
important
program of investigation of matter under extreme conditions that could
shed light
on the fundamental problems mentioned above.

In the present work we analyse the behavior of the neutron-proton
mass difference with temperature and density, within meson theory.
Our results are consistent with both partial chiral and charge  symmetry
restoration, with increasing temperature and density. In particular,
the density effects are in order to clear away the Nolen-Schiffer Anomaly.

Although a full description of the neutron-proton mass difference
calls for a non perturbative approach,  it is still lacking. Nevertheless,
one may gain a good
understanding of the problem, relying on the perturbative methods
used in many body nuclear physics. There, relativistic
perturbation theory (RPT) is mainly used for the analysis of hadron
interactions \cite{reps}.

In  a previous article  \cite{nos},  we have obtained a good outcome
for the n-p mass splitting using RPT at the hadronic level. In that work
we have shown that
the role of the rho-omega mixing interaction is crucial in the 
understanding
of the n-p mass difference, in a {\it hadronic context} (see also
\cite{gonz}).
We have concluded that the mixing of the vector
mesons coming from the u-d mass difference, has important consequences
on the nucleon self-energy and is the main non-electromagnetic
charge symmetry breaking (CSB) contribution to be considered.
This {\it effective} interaction has
been thoroughly investigated since 1960 \cite{gns}, particularly within
the tadpole picture \cite{mac}-\cite{lang}. It has been generally
claimed that the main source
for that vertex has a non-electromagnetic origin related to the quark
mass difference \cite{lang,pr90}. The connection between
the mixing lagrangian of eq.(1) and the microscopic quark level 
has been
recently explored within different models: QCD sum rules \cite{jin},
Coleman-Glashow tadpoles \cite{sca}, constituent quark 
model \cite{pr90}
and the Nambu--Jona-Lasinio model \cite{rein}.

Notice that $\Delta M^{exp}_{n-p}=1.29$ MeV while $\Delta
M^{em}_{n-p}=-0.66$ to $-0.76$ MeV \cite{cat,gas},
implying a 2 MeV strong contribution. Within the scheme of
ref.\cite{nos} this quantity can be obtained from the nucleon
self-energy given in eq.(2), using the experimental values
of couplings and meson masses in the literature
\cite{mac}-\cite{pr90},\cite{hm}.

\section{The model}

In the present work we obtain the finite temperature and density (FTD)
dependence of the
nucleon mass splitting using the same framework as in ref.\cite{nos}
where we have considered the nucleon and the vector mesons as the
fundamental dynamical degrees of freedom. In this context, the
calculation starts from the following Hamiltonian \cite{mac}
\begin{eqnarray}
{\cal H}_{I} & = & \frac{1}{2} g_{\rho} \bar{N}(p\prime)
                (\gamma^\mu +\frac{k^V}{2M} i \sigma^{\mu\nu} q_{\nu})
                \vec{\tau} \vec{\rho_{\mu}}(q) N(p)\nonumber\\
                & & + \frac{1}{2} g_\omega \bar{N}(p\prime)
                \gamma^\mu \omega_{\mu}(q) N(p) +
                \lambda \rho_{\mu}^0(q)\omega^{\mu}(q)
\end{eqnarray}

\noindent
which corresponds to the standard minimal formulation of the interaction
under consideration. Here $g_{\rho}$ and $g_{\omega}$ are the 
vector meson
coupling constants, $k^{V}$ is the isovector anomalous magnetic moment,
$M_n=M_p=M$,
$m_{\rho}$ and $m_{\omega}$ are the nucleon and mesons masses 
respectively
and $\lambda$ is the mixing matrix element.
The lowest order self-energy correction to the nucleon mass 
coming from
${\cal H}_{I}$, which explicitely breaks the charge symmetry is

\begin{eqnarray}
-i\Sigma^{F}(p) & = & \int \frac{d^4 q}{(2\pi)^4}\nonumber\\
        & & \left\{ \frac{g_\rho}{2}(\gamma^\mu +\frac{k^V}{2M} i
            \sigma^{\mu\nu} q_{\nu}) \tau_3 S(p-q)
         \frac{g_\omega}{2}\gamma_\mu \lambda D_{\rho}(q)
           D_{\omega}(q) + \right. \nonumber\\
        & & \left. \frac{g_\omega}{2}
\gamma^\mu S(p-q) \frac{g_\rho}{2}
(\gamma_\mu -\frac{k^V}{2M} i \sigma_{\mu\nu} q^{\nu})\tau_3
 \lambda D_{\omega}(q) D_{\rho}(q) \right\}
\end{eqnarray}
where
\vskip .5 cm

$ S(p-q)= \frac{(\not p-\not q +M)} {(p-q)^2-M^2}$\ \
and\ \ $ D_{v}(q)= 1 /(q^2-m_{v}^{2})$.

\vskip .5 cm \noindent
We will also include form factors at the meson-nucleon vertices
to take into account the effect of the nucleon structure in a
phenomenological way, thus giving to the model a wider range of
application in  its momentum dependence \cite{nos}

\[ F_{v}(q)= \frac{\Lambda_v^2-m_v^2}
{\Lambda_{v}^{2}-q^2}\hbox{\hspace{1cm}
where } v=\rho,\omega \]

A natural framework for the study of matter under FTD conditions 
is the
so-called Thermo Field Dynamics (TFD) \cite{tfd1}. TFD is a real 
time
formalism
of the
statistical field theory, very powerful for describing non-isolated
many body systems. It is a canonical field theory formulation in
which the Hilbert space is doubled and each field operator has two
independent components belonging to the thermal doublet. 
Correspondingly,
the Green's functions,
self-energies etc., are expressed by the thermal matrices. 
By virtue of this
extension of the Hilbert space, the pathologies of the pioneer 
formulations
of the real time formalism approach are avoided \cite{tfd2}. 
Moreover
because the Gell-Mann--Low formula
and the Wick's theorem for the perturbation expansion are 
available in TFD,
the usual perturbation theory at zero temperature and density 
can be easily
extended to FTD.
This formalism is also particularly useful to perform
both high and low temperature expansions,
 a feature much less accesible in the imaginary time 
formalism \cite{itf}.
Consequently, perturbation theory is at hand using the Feynman diagrams
technique proper of RPT. Moreover, the TFD free propagators can be
explicitely separated into two parts: one
term being the usual one and the other part depending on temperature
and density.
For the sake of brevity we will not discuss the derivation of the
corresponding thermal matrices for the modified propagators to be
used in the present calculation \cite{tfd3}. For fermions one gets

\vspace{0.5cm}
\begin{eqnarray}
S^{ab}_{\alpha\beta}(k) & = & S^{ab,F}_{\alpha\beta}(k)+
S^{ab,z}_{\alpha\beta}(k) \nonumber \\
                              & = & (\not\!k+M)_{\alpha\beta}(
\left( \begin{array}{cc}
         \frac{1}{k^2-M^2+i\epsilon} & 0 \\
          0 & \frac{1}{k^2-M^2-i\epsilon}
         \end{array} \right)   \nonumber \\
                              &   & \;\;\;\;\;+ \;
2\pi i \delta (k^2-M^2)
         \left( \begin{array}{cc}
         \sin^2(\theta_{k_{0}}) &  
\frac{1}{2} \sin(2\theta_{k_{0}}) \\
\frac{1}{2} \sin(2\theta_{k_{0}}) & -\sin^2(\theta_{k_{0}})
\end{array} \right) ) \;\;\;\;\;\;
\end{eqnarray}
\vspace{1 cm}
with
\begin{eqnarray}
\cos(\theta_{k_{0}}) & = & \frac{\Theta(k_{0})}{(1+e^{-z})^{1/2}}+
\frac{\Theta(-k_{0})}{(1+e^{z})^{1/2}} \nonumber \\
\sin(\theta_{k_{0}}) & = & \frac{e^{-z/2}
\Theta(k_{0})}{(1+e^{-z})^{1/2}}-
\frac{e^{z/2}\Theta(-k_{0})}{(1+e^{z})^{1/2}}
\end{eqnarray}

\vspace{1 cm} \noindent
where $z=(k_{0}-\mu)/k_{B}T$, \, $\mu$ is the relativistic chemical
potential
and $ \Theta(k_{0}) $ is the step function. Similarly for the meson
propagator one obtains

\begin{eqnarray}
D^{ab}(k) & = & D^{ab,F}(k)+ D^{ab,z}(k) \nonumber \vspace{.5 cm} \\
            & = & \left( \begin{array}{cc}
                         \frac{1}{k^2-m_{v}^2+i \epsilon} & 0 \\
                         0 & \frac{-1}{k^2-m_{v}^2-i \epsilon}
                         \end{array} \right)  \vspace{.5 cm} \\
            &   & \;\; - \; 2 \pi i \delta (k^2-m_{v}^2)
        \left( \begin{array}{cc}
         \sinh^2( \phi_{k_{0}}) &  \frac{1}{2} \sinh(2 \phi_{k_{0}})\\
        \frac{1}{2} \sinh(2 \phi_{k_{0}}) & -\sinh^2( \phi_{k_{0}})
         \end{array} \right) \nonumber
\end{eqnarray}

\vspace{0.5 cm}
\noindent
where the trigonometric functions for the particle-antiparticle
distributions have to be replaced by the corresponding hyperbolic
ones \cite{tfd3}.
The $ (a,b)=(1,1) $ elements of the thermal matrices are the standard
Feynman
physical propagators while the $(a,b)=(2,2)$ are the so-called
thermal ghost
partenaires and the off-diagonal terms are mixed in nature.
TFD establishes that inner vertices of a diagram can be of either type,
physical or ghost while the external ones ought to be physical.
In view of this, and up to second order perturbation, only the
mixing vertex $\lambda$ involve both kind of terms.
However, for the ghost type, the product of meson
propagators will produce a vanishing contribution to the loop.
Consequently, the full FTD expression for the second order self energy
results when $S^{11}$ replaces $S$ and $D_{v}^{11}$ replaces $D_{v}$
in eq.(2).
Now we can separate $\Sigma$ as
\vskip 0.5cm
\begin{eqnarray}
\Sigma & = & \Sigma^F + loop(S^{11,z}D_{\rho}^{11,F}D_{\omega}^{11,F})
                                        \nonumber \\
         &  & +loop(S^{11,F}D_{\rho}^{11,z}D_{\omega}^{11,F})
          + loop(S^{11,F}D_{\rho}^{11,F}D_{\omega}^{11,z})
\end{eqnarray}

\vspace{.5 cm} \noindent
where the first term, $\Sigma^F= loop(S^{F}D_{\rho}^{F}D_{\omega}^{F})$,
represents the usual zero temperature and density Feynman contribution
calculated in Ref.\cite{nos}, and the others are finite $T$-$\delta$
corrections.
On the other hand, for nucleons on-shell
$\delta M_p = {\bar u_{p}}(p) \Sigma u_{p}(p) =
- {\bar u_{n}}(p) \Sigma u_{n}(p)$, with $p^2=M^2$, then,
performing the $q_{0}$ integration, we obtain for the second
term in eq.(6) (see fig.(1))

\vspace{0.5 cm}
\begin{eqnarray}
 -i \delta^c M_p & = & -i \frac{\pi^2 g_{\rho} g_{\omega} \lambda}
 {(2 \pi M)^4}
 (\Lambda_{\rho}^2-m_{\rho}^2)(\Lambda_{\omega}^2-m_{\omega}^2)\tau_3
 \int_{M}^{\infty} d\mbox{$\varepsilon$}\ 
\frac{1}{e^{\beta( \mbox{$\varepsilon$}\ -\mu)}+1}
\label{tdc}\\
&  &  \frac{\sqrt{ \mbox{$\varepsilon$}^2\ -M^2} 
\,\, ((8+6k^V) M-(4+6k^V) \mbox{$\varepsilon$}\ )}
{(\frac{2M^2-m_{\rho}^2}{2M }-\mbox{$\varepsilon$}\ ) 
(\frac{2M^2-m_{\omega}^2}{2M}-\mbox{$\varepsilon$}\ )
(\frac{2M^2-\Lambda_{\rho}^2}{2M }-\mbox{$\varepsilon$}\ )
(\frac{2M^2-\Lambda_{\omega}^2}{2M }-\mbox{$\varepsilon$}\ ) }
\nonumber
\end{eqnarray}

\vspace{0.5cm}
\noindent
where the contribution coming from antiparticles has been omitted since
we will consider
temperatures much below pair production, $ k_B\beta ^{-1}<< M c^2$.
Furthermore,
as the presence of real mesons calls for extreme conditions, 
the in-medium 
corrections to the bosonic propagators included in the last two terms
of eq.(6), shall be neglected.

Since our description has been
given in terms of effective fields and couplings, it wouldn't make
sense to explore  our system in the high $T$-$\delta$ regime
where total deconfinement could take place. Hence it is natural 
to consider
$T<<M$ and densities not too high.
For arbitrary $T$ and $\delta$ conditions, eq.(\ref{tdc}) must be solved
numerically,
nevertheless one can find usefull expressions for some physical
scenarios. To this end one has to find accordingly, aproximated 
formulas
to the chemical potential which is defined in terms of 
temperature and
density by

\begin{equation}
\delta=\frac{4}{(2\pi\hbar)^3}\int d^3k \left( \frac{1}
{e^{(k_0-\mu)/k_BT}+1}-
\frac{1}{e^{(k_0+\mu)/k_BT}+1}\right)\label{defmu}
\end{equation}
At finite density, the chemical potential (in units $c=\hbar=k_{B}=1$)
can be aproximated by
\begin{equation}
\mu\simeq M+T \log \frac{\delta}{4} (\frac{2\pi}{MT})^{3/2}
\label{mudsmall}
\end{equation}
provided that \cite{landau}
\begin{equation}
2\pi(\frac{\delta}{4})^{2/3}<<TM<<M^2 \label{ht}
\end{equation}
Using eq.(\ref{mudsmall}) at low temperatures gives a very small
value for the self-mass correction $\delta^c M_p$ (see next section).
In particular, $\delta^c M_p$ vanishes as $T\rightarrow 0$.

On the other hand, in the range of the nuclear matter density,
a $T^2$ computation of $\mu$ results in
\begin{eqnarray}
\mu^2 & \simeq & \mu_{0}^2-\frac{\pi^2}{3}
(1+\frac{\mu_{0}^2}{\mu_{0}^2-M^2})T^2 \label{mudnuc}\\
\mu_{0}^2 & = & M^2+(\frac{6\pi^2}{4} \delta^{2/3})\nonumber
\end{eqnarray}
which is a good aproximation for
\begin{equation}
T^2<< \frac{\frac{9}{2}\delta^{4/3}}{(3\pi^2 \delta^{2/3}+M^2)}\label{lt}
\end{equation}
Now, performing a standard low T expansion \cite{kubo} in eq.(\ref{tdc}), 
we get
\newpage

\begin{eqnarray}
\delta^{c} M_p & \simeq & \frac{\pi^2 g_{\rho} g_{\omega} \lambda}
{(2 \pi M)^4}
(\Lambda_{\rho}^2-m_{\rho}^2)(\Lambda_{\omega}^2-m_{\omega}^2)
( \int_{M}^{\mu_{0}} g( \mbox{$\varepsilon$}\ ) 
d\mbox{$\varepsilon$}\ \nonumber\\
&  & +\frac{\pi^2}{6} T^2 (g'(\mu_0)-g(\mu_0) \frac{2\mu_{0}^2-M^2}
{(\mu_{0}^2-M^2) \mu_{0}^2}))
\label{corr}
\end{eqnarray}
\vspace{0.5 cm}


\noindent
with
\begin{equation}
g( \mbox{$\varepsilon$}\ )= 
\frac{\sqrt{ \mbox{$\varepsilon$}^2\ -M^2} \, 
((8+6k^V) M-(4+6k^V) \mbox{$\varepsilon$}\ )}
{(\frac{2M^2-m_{\rho}^2}{2M}- \mbox{$\varepsilon$}\ ) 
(\frac{2M^2-m_{\omega}^2}{2M}- \mbox{$\varepsilon$}\ )
(\frac{2M^2-\Lambda_{\rho}^2}{2M}-\mbox{$\varepsilon$}\ )
(\frac{2M^2-\Lambda_{\omega}^2}{2M}-\mbox{$\varepsilon$}\ )}
\label{g}
\end{equation}
\vspace{.3 cm}

\noindent
In this context, it can be seen from eq.(\ref{corr}), that $\delta^c
M_p$ is significant even at $T=0$.
Equations (\ref{mudnuc})-(\ref{corr}) are good aproximations
for astrophysical temperatures (a couple of MeV's) 
and baryon densities in
the range of the nuclear matter density $\delta_0=0.1934 fm^{-3}$
\cite{fw}.
Under these conditions,
the $T^2$-dependent term of eq.(\ref{corr}) is negligible.
Hence, the main contribution
comes from the first term, which only depends on $\delta$ 
(see next section).

Therefore, at finite density equations (\ref{mudsmall}) 
and (\ref{mudnuc})
have smooth limits for $T\rightarrow 0$, and so does eq.(\ref{tdc}).
On the other hand, at finite temperature and very low 
densities, one has
$\mu=\delta\cdot f(\beta)$,
implying that the limit $\delta\rightarrow 0$ is also smooth for the
self-mass correction (\ref{tdc}).
In this case, $\delta^{c} M_p$ depends on $e^{-M/T}$ 
and is exponentially
small for $T<<M$. The analysis above shows that
in the limit of vanishing chemical potential and zero
temperature, $\delta^c M_p\rightarrow 0$ as expected.

\vskip 1cm
\section{{\bf Numerical Results and Final Conclusions}}

In order to obtain numerical predictions from this approach we take
experimental values for all the physical quantities appearing
in our formulas:
$m_{\omega}=783$ MeV,$\  m_{\rho}=770$ MeV,$\  M=939$ MeV,$\  k^V=3.7$.
Concerning
the coupling constants we work with a set $g_{\rho} g_{\omega}
\lambda=$ 0.492 GeV$^2$  chosen in order to
saturate the 2 MeV hadronic contribution to the zero temperature
and density nucleon mass difference within the model.
This election is well inside the accepted range of variation of
these coupling constants found in the literature
\cite{mac,cb87,pr90,hm}.

As it is known, eq.(\ref{mudsmall}) is mostly suitable for high
temperatures and low densities, nevertheless
as our model is not expected to be reliable far from
deconfinement ($T_D\simeq 180$ MeV), we should consider $MT$ below
$T_D^2$ instead of $M^2$, together with conditions (\ref{ht}). For
example, at $T\leq 15$
MeV and densities satisfying these conditions 
(say below $\delta_0/10$),
the value of $\Delta^c M=\delta^c M_n-\delta^c M_p$ 
is below a couple of
keV, which is negligible (recall that in the chosen 
units, $1 fm=197$
MeV$^{-1}$). For low temperatures and baryon densities 
ranging from
$\delta_{0}/2$ to
$1.5\delta_{0}$, we have to use eqs.(\ref{mudnuc})-(\ref{corr}). 
In this
case, in turn, $\Delta^c M$ goes from
$-0.1$ to $-0.25$ MeV (see fig. 2).

Evidently, it is the non-trivial
shape  of $\mu(\delta,T)$ which is responsible for the behavior of the
in-medium mass difference.
As we have mentioned, the temperature contribution is negligible
for astrophysical temperatures, which we are interested in.

In connection with these results on the neutron-proton mass
difference as a function of the nuclear density,
let us finally comment on the Nolen-Schiffer Anomaly.
The NSA is a persistent problem in nuclear physics and the 
anomaly is the
failure of theory to explain the mass differences between 
mirror
nuclei ($i.e$ nuclei with $Z=A\pm 1/2$ and $N=A\mp 1/2$), 
a gap amounting
to a few hundred keV.
The effects of nuclear structure have been widely discussed 
with only
partial success \cite{slo}. The mass difference of mirror nuclei can
be written as $ M_{Z>}-M_{Z<}= \Delta E_{em}-(M_{n}-M_{p}) $
where $M_{Z>}$ is the mass of the nucleus with the larger
charge,
$\Delta E_{em}$ is the electromagnetic self energy difference
between the nuclei and
$M_{n}-M_{p}$ is the nucleon mass difference inside the nucleus.

Since $ \Delta E_{em} $ has been exhaustively analysed, 
in recent years
particular attention has been paid to the second term. 
A variety of models
have been put forward in order to avoid this problem. 
Generally
it is found  that $M_{n}-M_{p}$ is a decreasing function of the
nuclear density \cite{gonz}-\cite{krein}.
It means that within these models high density is also expected
to produce a partial charge symmetry restoration. This is the
expected behavior to deal with the anomaly.

In this way, the final
results of our calculation are in the right direction to 
remove the anomaly.

It should be mentioned that some recent literature has suggested
that the  $\lambda$ (off-shell) momentum dependence 
could be important
\cite{concha}. However, this conclusion
resulted from the use of oversimplyfied models for 
the $\rho-\omega$
correlation function. In fact, it has been recently 
shown \cite{jin}
that the effect of the different widths of the $\rho$ 
and $\omega$ mesons
implies  that $\lambda$ is almost momentum independent. 
This is consistent
with the present understanding of the CSB phenomenology
\cite{miller,thomas},
where the $\rho-\omega$ mixing has shown to be of 
particular importance.

Following similar steps to include the electromagnetic 
interaction
in the current scheme, we have found that the nuclear 
medium has a
relatively strong effect on the electromagnetic self-energy
$\Delta M_{n-p}^{\gamma}$ in the opposite direction,
ammounting to about $15 \% $
of its effect on $ \Delta M_{n-p}^{\rho \omega}$. This is
in good agreement with the results of Ref.\cite{gonz} which
have been derived making a quite
different treatment of the problem, in order to include the
external conditions by means of Skyrme type models.

In table 1 we show the values of NSA reported in the literature.
>From this table it emerges (a rough) accordance between the
predictions of the different models (with a not very clear density
dependent NSA).
For these nuclei
($\delta_{av.} \simeq \delta_0 /2$) our estimate of the NSA is
of 0.10 MeV. This is  consistent with the values quoted
in table 1 both {\it in sign and magnitude} although not big enough
to completely remove the anomaly.

These results for the n-p mass difference within dense
matter, suggest that in this case one should include, besides
the $\rho-\omega$ mixing, other contributions which, although
of minor importance in vacuum, seem to be relevant inside nuclei.

According to the vigor recently regained by the 
$\rho$-$\omega$ mixing
effective interaction \cite{jin}, in this work we have
extended our previous analysis on the mass splitting of an
isolated nucleon \cite{nos}. In the present article we have 
gone a step
further, considering the effects of temperature and density 
on this CSB
outcome.

For standard astrophysical temperatures, far from the
deconfining phase, we have found a negligible 
temperature effect on the
nucleon self-masses.
On the other hand, the density effects are significant
and were shown to produce the expected trends to 
remove the NSA.

\table{Table I. Values of the anomaly NSA in MeV reported
in Refs. \cite{nsa,slo,sato,9}. A is the mass number of mirror nuclei.}

\vspace{0.25 cm}
\begin{tabular}{||l|l|l|l|l||}\hline
  & { \scriptsize Nolen-Schiffer}  & { \scriptsize Shlomo} &
  {\scriptsize ~~~Sato} & {\scriptsize Hatsuda et al} \\
\ ~~A & ~~NSA\cite{nsa}  & NSA\cite{slo} & NSA\cite{sato}
& ~~NSA\cite{9} \\ \hline\hline
O-N~~~15 & ~~~-      & ~~0.16   & ~~0.29   & ~~0.53 \\ \hline
F-O~~~17 &  ~~0.31   & ~~0.31   & ~~0.19   & ~~0.29 \\ \hline
Si-P~~~29 & ~~0.20    & ~~~-     & ~~0.24   & ~~~-   \\ \hline
S-Cl~~~33 & ~~0.24    & ~~~-     & ~~0.28   & ~~~-   \\ \hline
Ca-K~~39 & ~~~-      & ~~0.22   & ~~0.43   & ~~0.57 \\ \hline
Sc-Ca 41 & ~~0.62    & ~~0.59   & ~~0.35   & ~~0.42 \\ \hline
\end{tabular}

\figure{Fig. 1. Diagrams contributing to the second term in eq. 6.
The broken line  represents the $T-\delta$ correction to the nucleon
propagator, $i.e: S^{11,z}$ (see eq. 3).}
\vskip 1cm
\figure{Fig. 2. Neutron-proton mass difference as a function of the
hadronic density $\delta$ in units of $\delta_0$, the nuclear matter
density, at T=0.}


\begin{references}

\bibitem{ft} S. Weinberg, Phys. Rev. {\bf D9} (1974), 3357.
           L. Dolan and R. Jackiw, Phys. Rev. {\bf D9} (1974), 3320.
\bibitem{latt} J. Kogut et al., Phys. Rev. Lett. {\bf 48} (1982), 1140.
L. Mc Lerran and B. Svetitsky, Phys. Lett. {\bf B98} (1981), 195.
         J. Polonyi et al., Phys. Rev. Lett. {\bf 53} (1984), 664.
\bibitem{eff}  N. Bilic, J. Cleymans and M. D. Scadron, Int. Jour. Mod.
Phys. {\bf A10} (1995) 1169; J. Cleymans, A. Kocic and M. D. Scadron,
Phys. Rev. {\bf C39} 323; D. Bailin, J. Cleymans and M. D. Scadron Phys.
Rev. {\bf C31} (1985) 164.
\bibitem{emc}  European Muon Collaboration,
                J. Aubert et al., Phys. Lett. {\bf B123} (1983), 275.
\bibitem{swel1} L. S. Celenza, A. Rosenthal and C. M. Shakin,
                Phys. Rev. Lett. {\bf 53} (1984), 892.
\bibitem{swel2} R. Jaffe, F. Close, R. Roberts and G. Ross,
                Phys. Lett. {\bf B134} (1984), 449.
\bibitem{nsa} J. A. Nolen, J. P. Schiffer, Annu. Rev. Nucl. Sci
                {\bf 19} (1969), 471.
\bibitem{gonz} L. N. Epele, H. Fanchiotti, C. A. Garcia Canal
                G. A. Gonzalez Sprinberg and R. Mendez Galain,
                Phys. Lett. {\bf B277} (1992), 33;
L. N. Epele, H. Fanchiotti, C. A. Garcia Canal and R.
                Mendez Galain, Phys. Rev. {\bf D39} (1989) R1473.

\bibitem{9}     T. Hatsuda, H. Hogaansen and M. Prakash,
                Phys. Rev. Lett. {\bf 66} (1991), 2851.
                E. Henley and G. Krein,
                Phys. Rev. Lett. {\bf 62} (1989), 2586.
\bibitem{krein} G. Krein, D. P. Menezes and M. Nielsen,
Phys. Lett. {\bf B294} (1992), 7.
L. A. Barreiro, A. P. Galeao and G. Krein, Phys. Lett.
{\bf B358} (1995) 7.
\bibitem{qmat} {\it Proceedings of Quark Matter '90}, Nucl. Phys.
{\bf A525} (1991), C. P. Singh, Phys. Rep. {\bf 236} (1993) 147.
\bibitem{reps}  K. Erkelenz, Phys. Rep. {\bf 13}, 191 (1974);
                K. Holinde, Phys. Rep. {\bf 68}, 121 (1981);
                R. Machleidt, K. Holinde and Ch. Elster,
                Phys. Rep. {\bf 149}, 1 (1987).
\bibitem{nos} H. R. Christiansen, L. N. Epele, H. Fanchiotti and
           C. A. Garcia Canal, Phys. Lett. {\bf B267}, 164 (1991).
\bibitem{gns} S. Glashow, Phys. Rev. Lett. {\bf 7} (1961) 469;
Y. Nambu and J. J. Sakurai, Phys. Rev. Lett. {\bf 8} (1962) 949.
\bibitem{mac}
P. C. Mc Namee, M. D. Scadron and S. A. Coon, Nucl. Phys. {\bf A249},
483 (1975); S. A. Coon, M. D. Scadron and P. C. Mc Namee, Nucl. Phys.
{\bf 287}, 381 (1977).
\bibitem{cb87} S. Coon and R. Barret, Phys. Rev. {\bf C36}, 2189 (1987).
\bibitem{lang} P. Langacker, Phys. Rev. {\bf D19} (1979) 2983;
P. Langacker and H. Pagels, Phys. Rev. {\bf D19} (1979) 2070.
\bibitem{pr90}
G. A. Miller, B. M. K. Nefkens and I. Slaus, Phys. Rep. 
{\bf 194} (1990) 1.
\bibitem{jin} M. J. Iqbal, X. Jin and D. B. Leinweber,
{\it Rho-omega mixing via QCD sum rules with finite mesonic widths},
TRI-PP-95-47, nucl-th/9507026.
\bibitem{sca} S. A. Coon and M. D. Scadron, Phys. Rev. {\bf C51} (1995)
2923.
\bibitem{rein} R. Friedrich and A. Reinhardt, {\it Rho-omega mixing
and the pion e-m form factors in the NJL model}. UNITU-THEP-1/1995,
hep-ph/9501333.

\bibitem{cat} M. Cini, E. Ferrari and R. Gatto, Phys. Rev. {\bf 2}, 7
(1959).
\bibitem{gas} J. Gasser and H. Leutwyler, Phys. Rep. {\bf 87}, 77
(1982).

\bibitem{hm}
E. M. Henley and G. A. Miller in {\it Mesons and Nuclei}, 
edited by M. Rho
and D. H. Wilkinson, (North Holland, Amsterdam, 1979) pag. 433;
A. G. Williams, A. W. Thomas and G. A. Miller, 
Phys. Rev. {\bf C36}, 1956 (1987);
M. Iqbal and J. Niskanen, Phys. Rev. {\bf C38}, 2259 (1988);
M. Beyer and A. G. Williams, Phys. Rev. {\bf C38}, 779 (1988);
M. Iqbal, J. Thaler and R. Woloshyn, Phys. Rev. {\bf C36}, 2442 (1987);
C. Cheung, E. M. Henley and G. A. Miller, Nucl. Phys. {\bf A305}, 342
(1978); {\it ibid.} {\bf 348}, 365 (1980);


\bibitem{tfd1} Y. Takahashi and H. Umezawa, Collect. Phen. {\bf 2}
(1975) 55; H. Matsumoto, Fortschr. Phys. {\bf 25} (1977) 1.
\bibitem{tfd2} A. J. Niemi, G. W. Semenoff, Ann. Phys. {\bf 152}
(1984) 105;
N. P. Landsman and Ch. G. van Weert, Phys. Rep. {\bf 145} (1987) 141.
\bibitem{itf} A major difficulty arising in the well-known
Matsubara formalism (ITF) is that one has to deal with Euclidean
propagators and Green functions with imaginary time arguments.
In principle, real time quantities can be obtained by analytic
continuation to the real axis but in practice the proper procedure
is not always self evident and may become a difficult task.
Nevertheless, at the one loop level, in can be proved that
the physical meaning of the different continuation procedures
are the same and the mentioned real quantities coincide with
those obtained in a real time formalism (See Landsman and
van Weert referenced above).
\bibitem{tfd3} K. Saito, T. Maruyama and K. Soutome Phys. Rev. {\bf C40}
(1989) 407.
\bibitem{kubo} R. Kubo in {\it Statistical Mechanics} North Holland
Pub.Co. Amsterdam, 1965.
\bibitem{landau}
L. Landau, E. Lifchitz and L. Pitayevski, {\it Physique
Statistique}, Editions Mir, Moscou, 1967 (see ch.5 and \S 27,45.)
\bibitem{fw} A. L. Fetter, J. D. Wallecka in
{\it Quantum Theory of Many-Particle Systems}, Mc Graw-Hill Pub.Co.
1971.
\bibitem{slo} S. Shlomo, Rep. Progr. Phys. {\bf 41} (1978), 957.
\bibitem{adami} We have used the vacuum values for masses and couplings.
The deviation from these numbers, a higher order correction, is
negligible at low temperatures and densities. See e.g. C. Adami and
G. E. Brown, Phys. Rep. {\bf 234} (1993) 1; S. Gao, R-K. Su and
P. K. N. Yu, Phys. Rev. {\bf C49} (1994) 40, and references therein.
\bibitem{sato} H. Sato, Nucl. Phys. {\bf A269} (1976) 378.

\bibitem{concha} J. Piekarewicz and A. G. Williams, Phys. Rev. {\bf C47}
(1993) R2462; H. B. O'Connell, B. C. Pierce, A. W. Thomas and A. G.
Williams, Phys. Lett {\bf B336} (1994) 1.

\bibitem{miller} G. A. Miller and W. T. H. van Oers,
{\it Charge Independence and Charge Symmetry}, nucl-th/9409013.
\bibitem{thomas} A. W. Thomas and K. Saito, {\it Charge Symmetry
Violation in Nuclear Physics}, ADP-95-37/T191.

\end{references}
\end{document}